\documentclass{ceab}   
\usepackage{color}
\usepackage[dvipsnames]{xcolor}
\usepackage{epsfig}     
\usepackage{graphicx}   
\usepackage{url}
\usepackage{natbib}     
\usepackage[T1]{fontenc} 
\usepackage{babel}       

\usepackage{hyphenat}
\def\runningtitle{J.C: Arteaga-Velázquez: Cosmic rays from 10 TeV to 1 EeV}   
\def\articlenumber{7}
\def\ppages{1--4}

\begin{document}

\title{Recent measurements of the spectrum and composition of cosmic rays between
$10$ TeV and $1$ EeV from EAS experiments}

\author{J.C. Arteaga-Vel\'azquez$^{1}$\thanks{email: juan.arteaga@umich.mx}
\vspace{2mm}\\
\it $^1$Instituto de F\'\i sica y Matem\'aticas, Universidad Michoacana, Morelia,\\ 
\it Michoacan, Mexico\\
}

\maketitle

\begin{abstract}
\scriptsize
In the last two decades, different advances with extensive air shower (EAS) experiments have allowed to unveil several
details about the composition and energy spectrum of galactic cosmic rays from some TeV up to several hundreds of PeV, which have led to improve our understanding of the physics of cosmic rays. EAS measurements above the knee, for instance, have helped to understand some aspects about the acceleration and propagation mechanisms of the PeV component of cosmic rays and have shown possible hints about the transition from the galactic to the extragalactic regime. Meanwhile, EAS data in the TeV energy range have revealed unexpected features in the all-particle and elemental spectra, which may indicate the presence of previously unknown cosmic-ray phenomena. In this contribution, I will present a brief account on these EAS results, with particular emphasis on the recent measurements of the all-particle energy spectrum and the elemental com\-po\-sition of cosmic rays from $10$ TeV up to $1$ EeV. In addition, I will compare the latest results of EAS experiments and direct
detectors at energies just below the knee, where there exists an overlap between both detection
techniques. 
\end{abstract}

\keywords{Cosmic rays, extensive air showers, cosmic ray experiments, cosmic ray energy spectrum,
cosmic ray composition}

\section{Introduction}

 The discovery of cosmic rays by V. F. Hess in 1912 opened a new astrophysical window to the sky and revealed the existence of enigmatic high-energy physics phenomena in the universe (\textcolor{cyan}{Hess, 1912}).The maximum energy achievable in such processes was not realized at that moment due to the lack of dedicated cosmic-ray measurements. However, with the passing of time and the development of new detection techniques, it was discovered that cosmic rays can bear energies several orders of magnitude beyond the ones observed in particle beams of radioactive decays and current man-made accelerators.  Early measurements of the energy spectrum around 1938 using electroscopes at the top of the atmosphere in different geomagnetic latitudes revealed cosmic-ray particles with energies from $1.4$ GeV up to $10$ GeV (\textcolor{cyan}{Bowen et al., 1938}). Then, in 1939, P. Auger's data on EAS showed the existence of cosmic rays with energies of $\sim 1$ PeV (\textcolor{cyan}{Auger et al., 1939}).  The presence of primary particles with energies close to $10$ PeV was observed at the beginning of the $50$'s with the Cornell array (\textcolor{cyan}{Barret et al., 1952}). At the middle of the same decade, it appeared evidence in favor of the arrival of cosmic rays with energies of $100$ PeV (\textcolor{cyan}{Crawshaw et al., 1956}). In 1957, the MIT EAS experiment (\textcolor{cyan}{Clark et al., 1957}) found that  cosmic rays  can reach energies even of  $1$ EeV and by 1963, with the Volcano Ranch array, J. Linsley  demonstrated the existence of cosmic rays with energies as high as $10^{20}$ eV (\textcolor{cyan}{Linsley, 1963}). At the moment, the highest primary energy measured in a cosmic-ray particle is $3.2 \times 10^{20}$ eV, and it was reported by the Fly's Eye Collaboration (\textcolor{cyan}{Bird et al., 1995}). This makes of cosmic rays the most energetic form of extraterrestrial radiation. How and where this radiation is produced, and how it crosses the universe when traveling to the Earth are questions that do not have clear answers yet.  One of the objectives of cosmic-ray physics is to provide measurements on the energy spectrum and composition of cosmic rays to provide some clues that could help to solve this enigma. 

  To measure cosmic rays below $10^{15}$ eV, direct techniques are employed using particle detectors, such as calorimeters, nuclear emulsions, spectrometers, etc., on board of balloon flights, planes, spacecrafts and satellites (\textcolor{cyan}{Maestro, 2015}). Meanwhile, at energies greater than $10^{13}$ eV, they are studied by indirect methods using EAS arrays composed of particle detectors, radio antennas, calorimeters, Cherenkov and fluorescence telescopes (\textcolor{cyan}{Haungs et al., 2003}). There is a common energy region for the study of cosmic rays with both methods, which is located between $10^{13}$ and $10^{15}$ eV. This particular energy range offers the opportunity to perform comparative studies between the direct and indirect methods of detection, which could help to perform cross calibrations and studies of systematic errors. 

  From cosmic ray measurements, it is known that the energy spectrum of these particles approximately follows a power-law behavior $\propto E^{\gamma}$, where $\gamma$, the spectral index, is close to $-3$ (\textcolor{cyan}{Mollerach et al., 2018}). This way the intensity rapidly decreases with the primary energy. This generally constrains direct measurements to energies below $1$ PeV, as the detectors employed in these studies have limitations on size. EAS detectors, on the contrary, can cover large surfaces and, hence, compensate for such effect. As a consequence, they allow to have large statistics at energies above $1$ PeV.

\begin{figure}[ht!]
   \centering
   \includegraphics[scale=0.5]{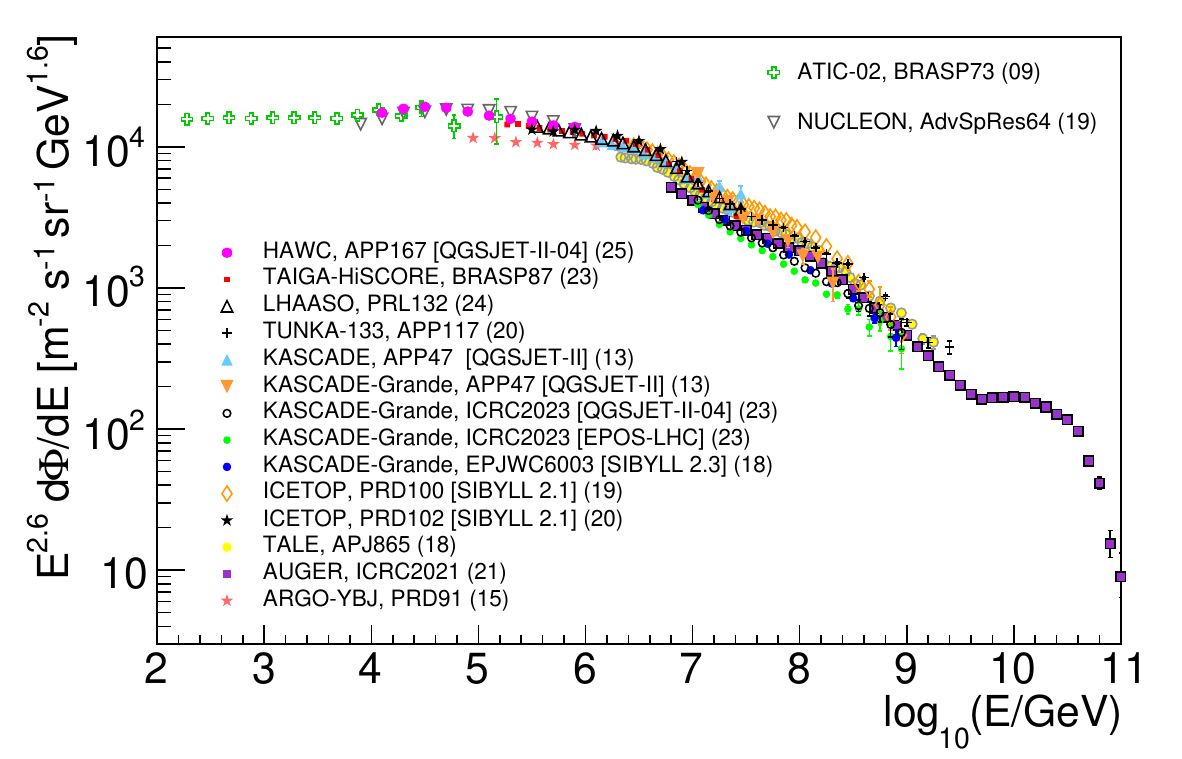}%
   \vspace{-1.3pc}
  \caption{The all-particle energy spectrum of cosmic rays according to recent measurements from different experiments above $100$ GeV.}
  \label{Allspec}
  \end{figure}

  Air shower studies  have allowed to identify three main structures in the spectrum of cosmic rays (\textcolor{cyan}{Mollerach et al., 2018}), see Fig.~\ref{Allspec}. The first one is a softening at around $4$ PeV, where the spectral index decreases from $-2.7$ to $-3$. This feature is called the {\it knee} and  was discovered with the MSU EAS array (\textcolor{cyan}{Kulikov et al., 1958}). The second feature is a hardening close to $5$ EeV that is called the {\it ankle}, which is characterized by an increment in $\gamma$ from $-3.3$ to $-2.6$. It was first hinted by the Volcano Ranch experiment (\textcolor{cyan}{Linsley, 1963a}) and later confirmed with the Haverah Park array (\textcolor{cyan}{Cunningham et al., 1980}). The last feature is a suppression at $\sim 50$ EeV, where the spectral index drops to $\sim -5.2$ (\textcolor{cyan}{Abreu et al., 2021}). Clues about the existence of this cutoff were devised in (\textcolor{cyan}{Bahcall and Waxman, 2002}) from Yakutsk (\textcolor{cyan}{Efimov et al., 1991}), Fly's Eye  (\textcolor{cyan}{Bird et al., 1993; Bird et al., 1994}) and HiRes (\textcolor{cyan}{Abu-Zayyad et al., 2002}) data. However, the suppression was definitely proved by the HiRes experiment (\textcolor{cyan}{Abbasi et al., 2008}) and the Pierre Auger observatory (\textcolor{cyan}{Abraham et al., 2008}).
  
  There exists also  secondary features in the energy spectrum, which have been discovered in the last twelve years (\textcolor{cyan}{Alfaro et al., 2017}; \textcolor{cyan}{Mollerach et al, 2018}; \textcolor{cyan}{Castellina et al., 2023}). In particular, a hardening at $20$ PeV, which is known as the {\it low-energy ankle}, a  {\it second knee}, which is observed at $\sim 100$ PeV, a  cutoff at around tens of TeV, and another softening at $13$ EeV, named as the {\it instep}. The first structure is generated by an increment in the spectral index up to $-2.9$ after the fall-off of the spectrum at the knee, and the second one\footnote{\scriptsize The second knee was originally referred to a steepening of the total spectrum in the $4-7$ PeV interval (\textcolor{cyan}{Bergman et al., 2007}) that was first pointed out by Haverah Park (\textcolor{cyan}{Lawrence et al., 1991}) and AKENO (\textcolor{cyan}{Nagano et al., 1992}). It is not clear whether this feature is different from the one at $100$ PeV yet. \vspace{-1pc}}, by a reduction to $\gamma = -3.3$, after the recovery at the low-energy ankle. Both of them were reported by the KASCADE-Grande Collaboration (\textcolor{cyan}{Apel et al., 2012}). The TeV softening was observed with the HAWC observatory in 2017 (\textcolor{cyan}{Alfaro et al., 2017}), but some clues about its existence were previously presented by the NUCLEON satellite experiment in the same year (\textcolor{cyan}{Atkin et al., 2017}). According to (\textcolor{cyan}{Alfaro et al., 2025}), the spectral index changes from  $-2.5$ to $-2.7$ at the TeV softening.  Finally, the instep was recently discovered by the Auger Collaboration (\textcolor{cyan}{Aab et al. 2020}). Here, the spectrum changes from $\gamma = -2.5$ to $-3.1$.

\begin{figure}[t!]
   \centering
   \includegraphics[scale=0.5]{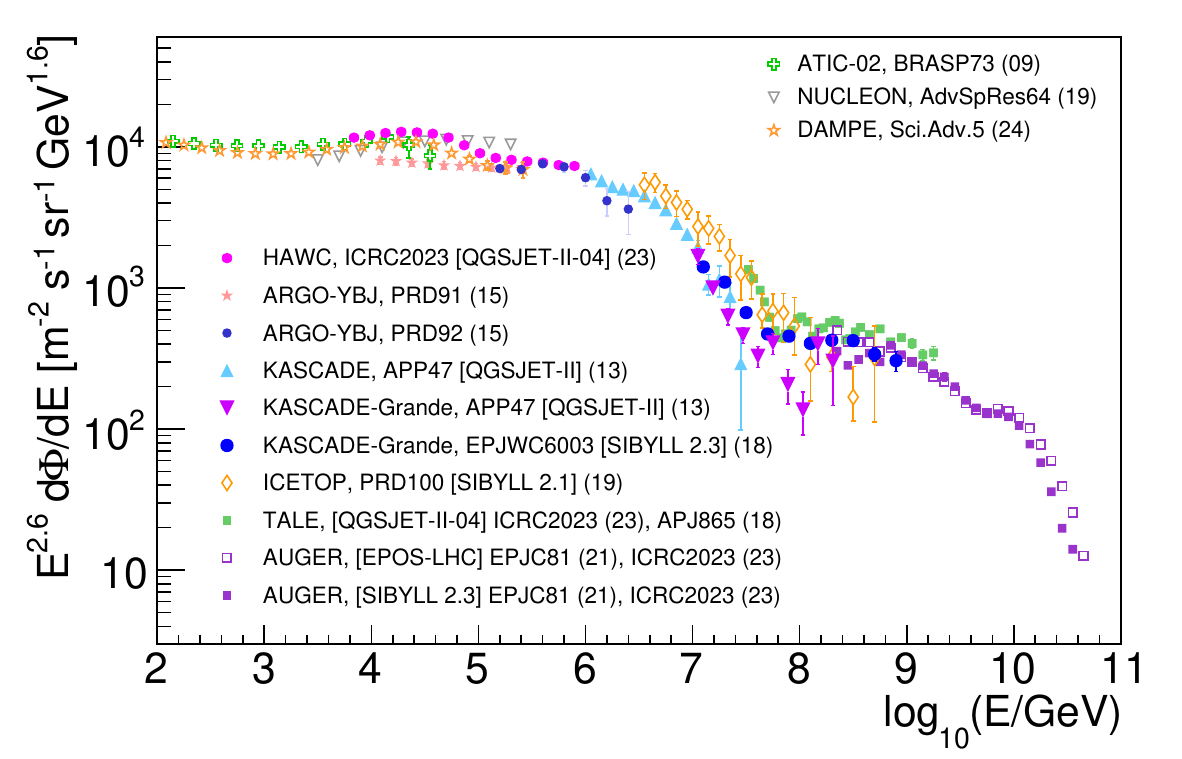}\\
   \includegraphics[scale=0.5]{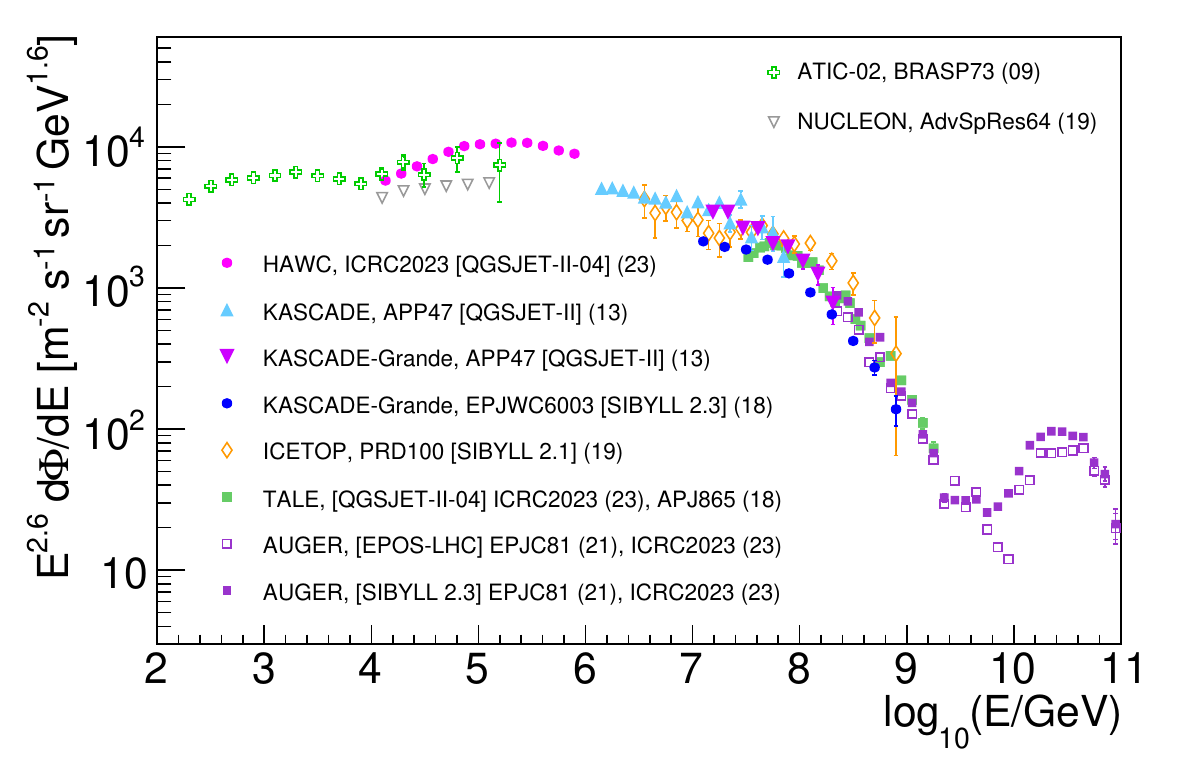}
   \vspace{-1.3pc}
    \caption{Energy spectra for the light (H$+$He) and heavy ($Z \geq 3$) mass groups of cosmic rays from various experiments between $100$ GeV and $100$ EeV (top and bottom panels, respectively). For the Auger spectra, the fractions of cosmic ray nuclei  from  (\textcolor{cyan}{Tkachenko et al., 2023}) were used on the total spectrum of  (\textcolor{cyan}{Abreu et al., 2021}), and for TALE, the fractions from (\textcolor{cyan}{Fujita et al., 2023}) on the spectrum of (\textcolor{cyan}{Abbasi et al., 2018}).}
   \label{Light-Heavy}
  \end{figure}

\begin{figure}[h!]
   \centering
   \includegraphics[scale=0.48]{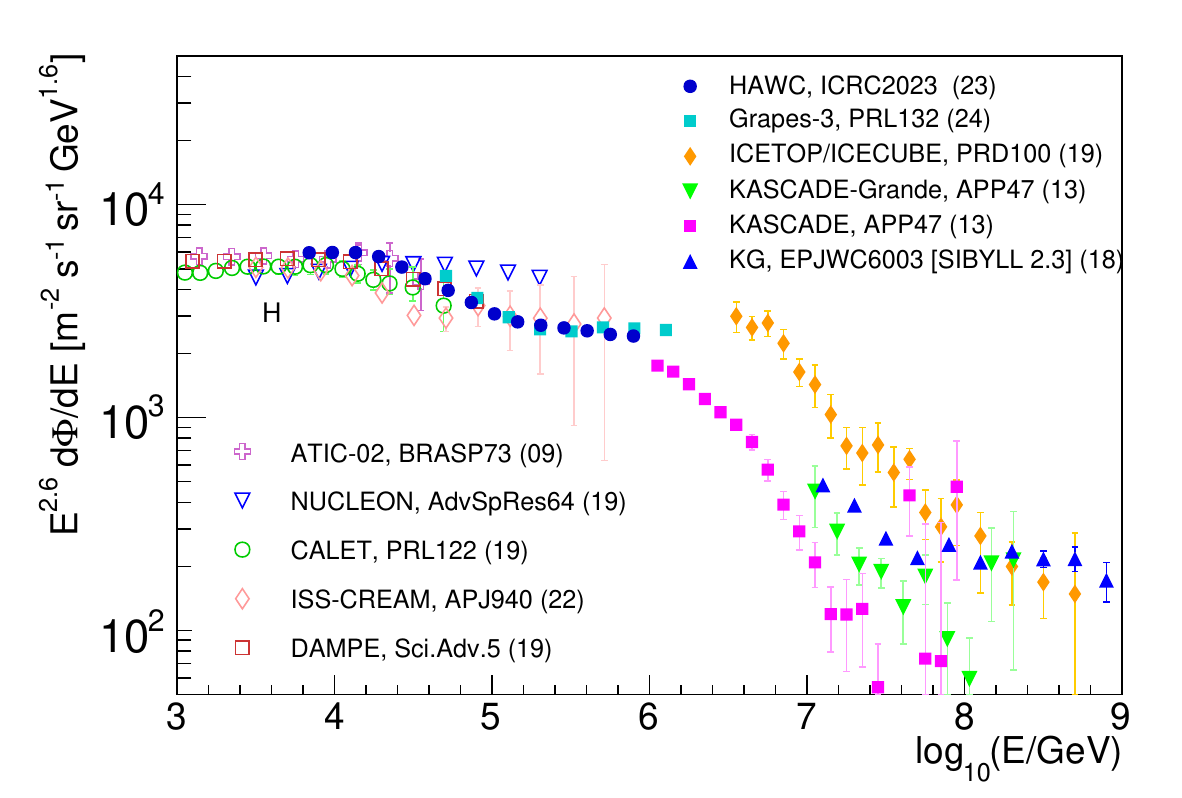}    
   \includegraphics[scale=0.48]{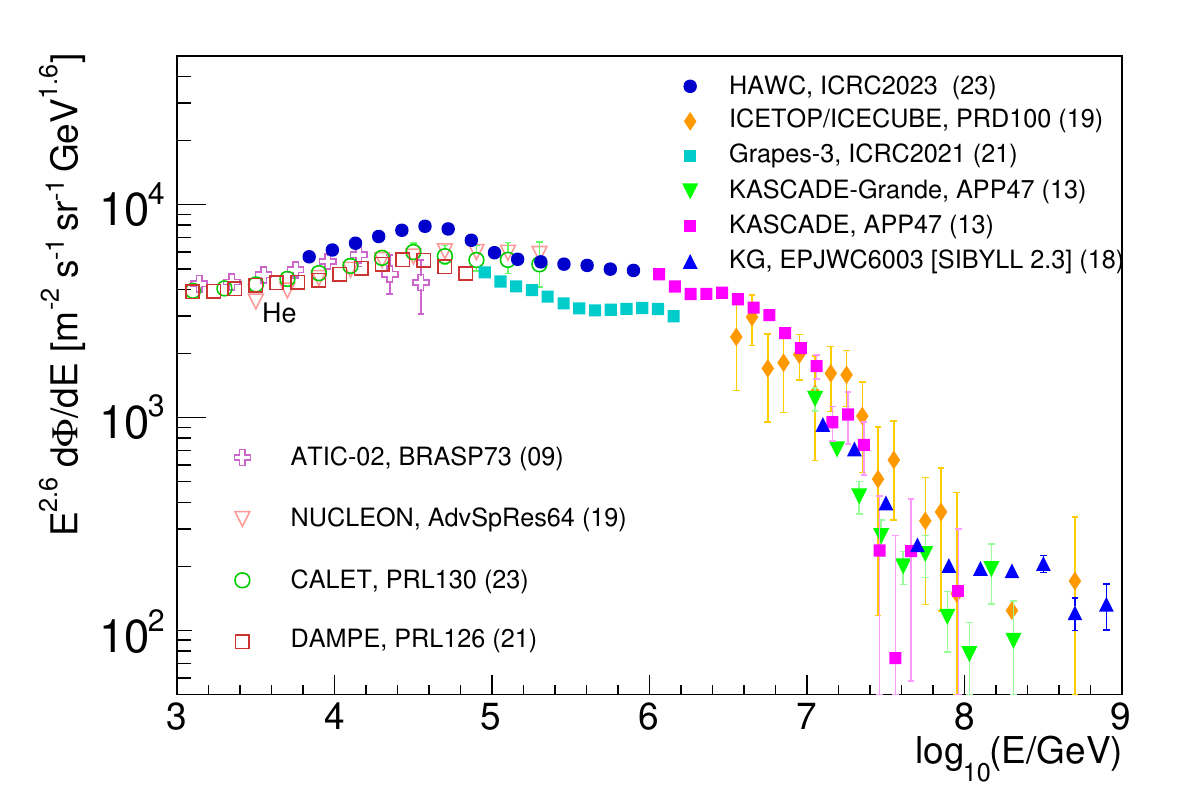}
   \vspace{-1.3pc}
    \caption{The energy spectrum for H and He primaries measured with various detectors between $1$ TeV and $1$ EeV (top and bottom plots, respectively).}
    \label{H-He}
  \end{figure}

  Investigations of the composition of cosmic rays have shown that at the energies at which the features in the all-particle spectrum are located there are important changes in the relative abundances of the primary particles (\textcolor{cyan}{Mollerach et al, 2018}), c.f. Fig.~\ref{Light-Heavy}. Cosmic rays are mainly composed by atomic nuclei with a small contribution from electrons, positrons, neutrons and light antinuclei (\textcolor{cyan}{Navas et al., 2024}). The most abundant components in the cosmic-ray flux are protons and helium. However, they do not dominate the spectrum of cosmic rays in the full energy range, as their relative abundances are energy dependent, c.f. Fig.~\ref{H-He}. Cosmic ray data show that the light nuclei of cosmic rays (H and He) are more abundant for energies below $10$ PeV (\textcolor{cyan}{Mollerach et al, 2018}) and in the interval $1 - 10$ EeV (\textcolor{cyan}{Mayotte et al., 2023}). Their energy spectrum exhibit a steepening at around $24$ TeV (\textcolor{cyan}{Albert et al., 2022}; \textcolor{cyan}{Arteaga et al., 2023}; \textcolor{cyan}{Alemanno et al., 2024}) and a cutoff at $\sim 4$ PeV (\textcolor{cyan}{Antoni et al., 2005}), which are located around the positions of the TeV softening and the knee in the all-particle energy spectrum, respectively. Meanwhile at $1$ and $10$ EeV,  there seems to exist two additional steepenings in the spectrum of H+He, respectively, as suggested by the measurements of the Pierre Auger observatory (\textcolor{cyan}{Mayotte et al., 2023}). Here, the second spectral feature, at  $10$ EeV, seems to occur at the position of the instep.  With regard to the heavy component, its energy spectrum  has a steepening at $100$ PeV, which causes the second knee in the all-particle spectrum, while a transition from the H$+$He$+$C component to the Si$+$Fe mass group at $10$ PeV produces the low-energy ankle (\textcolor{cyan}{Apel et al., 2013}). The suppression at ultra-high energies in the total spectrum seems to be originated in a cutoff in the spectrum of heavy nuclei (\textcolor{cyan}{Mayotte et al., 2023}).

  The features in the energy spectrum of cosmic rays and the particular evolution of the composition of these particles are the result of the production mechanisms and the physical processes that affect the propagation of cosmic rays in different energy regimes. At energies below $100$ PeV or $1$ EeV, cosmic rays are of galactic origin, but at higher energies, they are produced beyond the Milky Way (\textcolor{cyan}{Aab et al., 2017; Mollerach et al, 2018}). A galactic-extragalactic transition is expected in the interval of $100 \, \mbox{PeV} - 1 \, \mbox{EeV}$ (\textcolor{cyan}{Kachelriess, 2019}). Cosmic rays may be accelerated through the first-order Fermi's acceleration mechanism, which involves multiple interactions with the front of astrophysical shocks and magnetic confinement at the source  (\textcolor{cyan}{Ostrowski et al, 2002}). Recently, however, it has been also argued that  magnetically dominated turbulence at the sources could also  accelerate cosmic rays up to ultra-high energies (\textcolor{cyan}{Comisso et al., 2024}).
  The propagation through the space seems to occur in a diffusive way due to interactions with the the extraterrestrial  magnetic fields and their fluctuations  (\textcolor{cyan}{Giacinti et al., 2023}).  There is a maximum energy at which magnetic fields can not contain anymore cosmic rays, therefore they start to escape from the magnetized region producing a cutoff in the elemental spectra of the primary particles. The energy at which it happens is proportional to the charge of the nuclei, the intensity of the magnetic field and the size of the confinement region. This is what is called the Peter's cycle  (\textcolor{cyan}{Peters, 1961}).  

  The Peter's cycle seems to be responsible for the knee and the second knee, as pointed out by the KASCADE-Grande data (\textcolor{cyan}{Apel et al., 2013}). This process would occur inside the galactic  sources or in the interstellar magnetic field. The escaping of the light and heavy nuclei from these regions would lead to the knee and  second knee, respectively. At lower energies, there are also  hints provided by NUCLEON, which would indicate that the TeV softening could have an origin in a Peter's cycle (\textcolor{cyan}{Atkin et al., 2017}). At ultra-high energies this scenario may be plausible (\textcolor{cyan}{Abdul et al., 2024}). More research and data is needed to test this possibility at TeV and EeV energies. In the following sections, a selection of recent measurements on the composition and  energy spectra of cosmic rays in the $10$ TeV - $1$ EeV energy range from EAS experiments is presented. The review is not intended to be exhaustive and it reflects the personal perspective of the author. 

    \section{The $10$ TeV$-$$1$ PeV region}

     In this energy regime, the ARGO-YBJ experiment ($2007 - 2013$) measured  the all-particle energy spectrum in the energy interval from $70$ TeV to $20$ PeV and  the spectrum of light nuclei (H$+$He) between $3$ TeV and $3$ PeV with large statistics (\textcolor{cyan}{Montini et al., 2016}). The experiment found a cutoff at $700$ TeV in the spectrum of H$+$He, implying that the cosmic-ray composition becomes heavy towards the knee. At lower energies, ARGO-YBJ reported a plain power-law behavior for the spectrum of light nuclei. 

     Recently, the HAWC collaboration derived the all-particle spectrum between $10$ TeV and  $1$ PeV, thus filling the gap that appears at the frontier between direct and indirect experiments with high statistics shower data (\textcolor{cyan}{Alfaro et al., 2025}). In addition, it confirmed the presence of the TeV softening in the total spectrum, which was observed at $\sim 40$ TeV, and performed comparative studies with direct measurements, which showed that the HAWC spectrum is in good agreement with NUCLEON (\textcolor{cyan}{Grebenyuk et al. 2019}) and ATIC-02  (\textcolor{cyan}{Panov et al., 2009}) data above $10$ TeV. In another analysis,  HAWC  estimated the spectrum of H$+$He nuclei between $6$ and $158$ TeV and reported a knee-like feature at $24$ TeV (\textcolor{cyan}{Albert et al., 2022}). The existence of this feature is in agreement with recent results from the DAMPE telescope (\textcolor{cyan}{Alemanno et al., 2024}), but it does not support ARGO-YBJ observations about a featureless spectrum in this energy interval. A further study of the HAWC collaboration has allowed to estimate the cosmic-ray spectra for H, He and the heavy mass group ($Z \geq 3$) from $10$ TeV to $1$ PeV revealing the presence of several spectral features (\textcolor{cyan}{Arteaga et al., 2023}). In particular, it observed the softenings in the H and He spectra at tens of TeV, which were previously reported by DAMPE (\textcolor{cyan}{An et al., 2019}; \textcolor{cyan}{Alemanno et al., 2021}), CALET (\textcolor{cyan}{Adriani et al. 2019}; \textcolor{cyan}{Adriani et al. 2023}), NUCLEON (\textcolor{cyan}{Grebenyuk et al. 2019}) and ISS-CREAM (\textcolor{cyan}{Choi et al., 2022}), and showed the existence of another softening, albeit in the spectra of heavy nuclei at around $300$ TeV. It also found that the steepening detected at TeV energies in the spectrum of H$+$He (\textcolor{cyan}{Albert et al., 2022}) is generated by the superposition of the individual breaks observed in the spectra of protons and helium nuclei, and that the TeV softening in the total spectrum seems to be produced mainly by the cutoff in the elemental spectrum of light primaries. Furthermore, it pointed out the presence of individual hardenings in the spectra of H and He at $\sim 100$ TeV. HAWC results on composition also showed that He nuclei are more abundant than protons in this energy regime, and that the relative abundance of the heavy component increases at energies from $10$ to $300$ TeV and decreases towards $1$ PeV.

     The  hardening in the spectrum of protons has been also reported by the GRAPES-3 experiment, in particular, at energies of $160$ TeV (\textcolor{cyan}{Varsi et al., 2024}). GRAPES-3 has also published preliminary studies that support the existence of a similar feature in the spectrum of He nuclei at around $400$ TeV (\textcolor{cyan}{Varsi et al., 2021}). Both structures have not been confirmed by direct experiments yet, however, there are clues about such kinks in the spectra of H$+$He nuclei and protons in the DAMPE (\textcolor{cyan}{Alemanno et al., 2024}) and ISS-CREAM data (\textcolor{cyan}{Choi et al., 2022}), respectively. 

     LHAASO-KM2A has just started to publish results on its analyses on cosmic rays. In (\textcolor{cyan}{Cao et al., 2024}), it was derived the all-particle energy spectrum in the $300$ TeV$- 30$ PeV interval with the higher accuracy to date and probed the composition of cosmic rays by estimating the mean logarithmic mass of the primary nuclei $\langle \ln A \rangle$. The results of LHAASO-KM2A did not support the existence of an increment of the abundance of the heavy component above $700$ TeV as suggested by ARGO-YBJ. 

     In the last years, ICETOP succeeded in extending its measurements of the total energy spectrum from the PeV region towards the $100$ TeV regime, by providing data  in the interval of $250 \, \mbox{TeV} - 10 \, \mbox{PeV}$ (\textcolor{cyan}{Aartsen et al., 2020}). Furthermore, TAIGA-HiSCORE updated its results on the total spectrum of cosmic rays, which extends from $220 \, \mbox{TeV}$ up to $28  \, \mbox{PeV}$  (\textcolor{cyan}{Prosin et al., 2023}). In general, below $1$ PeV, the results of HAWC, LHAASO-KM2A, ICETOP and TAIGA-HiSCORE for the all-particle spectrum are in good agreement within systematics uncertainties, however, they are higher than the ARGO-YBJ data  in the same energy range as seen in Fig.~\ref{Allspec}.

    \section{The $1$ PeV$-$$1$ EeV regime}

     The KASCADE (1996-2013) and KASCADE-Grande (2003-2013) experiments have contributed in an important way to the study of the spectrum and composition of cosmic rays in the energy ranges of $1 - 100$ PeV and $10 \, \mbox{PeV} - 1 \, \mbox{EeV}$, respectively (\textcolor{cyan}{Arteaga et al., 2015}). KASCADE measured the position of the knee at energies between $4$ and $6$ PeV and showed that the total spectrum at these energies is dominated by the elemental groups of H, He and C, whose intensities were unfolded from EAS data on the electron and muon sizes. It also found out that the spectra of the light (H, He) and medium (C$-$Si) mass groups of cosmic rays have individual cutoffs (\textcolor{cyan}{Antoni et al., 2005}; \textcolor{cyan}{Apel et al., 2013}). Even more, KASCADE determined that the break energies for these structures increase with the atomic number of the primary nuclei. In particular, using QGSJET-II  (\textcolor{cyan}{Ostapchenko, 2006}) to interpret the data, the features were observed at energies of  $4$ PeV for protons, $7$ PeV for He and $20$ PeV for C. No evidence was found in the KASCADE data about the existence of a kink in the spectrum of Fe nuclei. 
     
\begin{figure}[t!]
   \centering
   \includegraphics[scale=0.4]{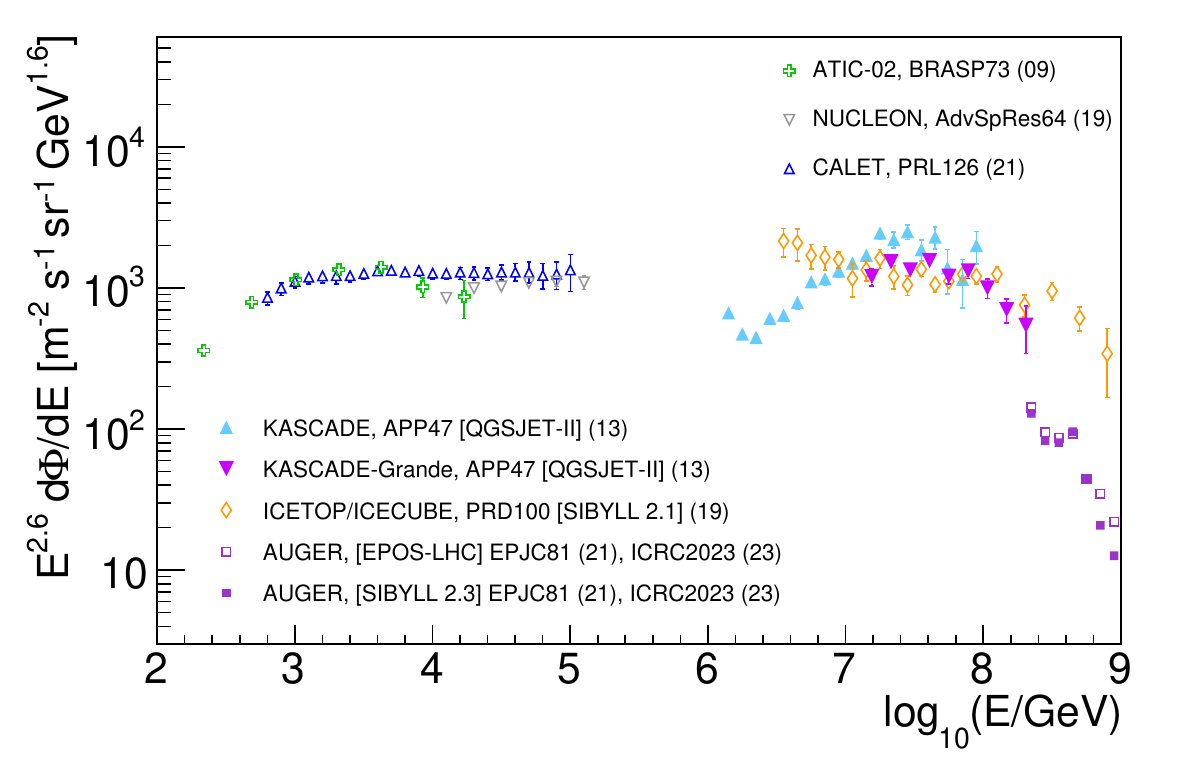}
   \vspace{-1.3pc}
  \caption{The spectrum of iron nuclei from recent cosmic-ray experiments in the interval from $100$ GeV to $1$ EeV.  For the Auger spectra, the fractions of cosmic ray nuclei  from  (\textcolor{cyan}{Tkachenko et al., 2023}) were used on the total spectrum of  (\textcolor{cyan}{Abreu et al., 2021}).}
  \label{Fespectrum}
  \end{figure}

     KASCADE-Grande extended the measurements of KASCADE to higher energies and unfolded the  elemental spectra of H, He, C, Si and Fe nuclei between $10$ PeV and $1$ EeV from the number of charged particles and muons in EAS (\textcolor{cyan}{Apel et al., 2013}). From this data, a knee-like feature in the spectrum for Fe primaries at around $80$ PeV was discovered, see Fig.~\ref{Fespectrum}. The  analyses of the KASCADE-Grande collaboration of the energy location for the knee-like features of the spectra of H, He, C and Fe nuclei, which are observed in the data of KASCADE and KASCADE-Grande, have demonstrated that they are rigidity dependent (\textcolor{cyan}{Apel et al., 2013}). Albeit, the steepening of the spectrum for silicon does not fit to this picture. This, however, may arguably be the result of an incompatibility  between the data and the high-energy hadronic interaction models used in these investigations. Here, it is important to mention that the KASCADE-Grande collaboration has shown that the presence of features in the elemental energy spectra are independent of the low- and high-energy hadronic interaction models used in the unfolding procedure, although the relative abundances are model dependent (\textcolor{cyan}{Antoni et al., 2005}; \textcolor{cyan}{Apel et al., 2009}).

     The data from  KASCADE-Grande on the energy spectra of the cosmic-ray mass groups show that the primary composition becomes heavier towards $100$ PeV, in the region of the second knee, and that this feature is produced from the cutoff in the iron spectrum (\textcolor{cyan}{Fuhrmann et al., 2012}; \textcolor{cyan}{Apel et al., 2013}). The connection of the heavy component of cosmic rays with the second knee was also independently established in a data-driven analysis carried out in (\textcolor{cyan}{Apel et al., 2011}) by KASCADE-Grande. In this study, the energy spectrum of electron-poor showers, dominated  by heavy primaries (Si$+$Fe), was extracted from the KASCADE-Grande data based on cuts on a sensitive parameter estimated from the shower and muon sizes. The result showed that a knee-like feature close to $80$ PeV in the intensity of this component causes the second knee. In a similar analysis, the energy spectrum for electron-rich events, which contain mostly H and He nuclei, was reconstructed with KASCADE-Grande data, revealing an ankle-like feature in the light component of cosmic rays at around $120$ PeV produced by a change of spectral index from $-3.25$ to $-2.79$ (\textcolor{cyan}{Apel et al., 2013a}). Both studies have been recently updated  in (\textcolor{cyan}{Kang et al., 2023}) using the post-LHC hadronic interaction models QGSJET-II-04 (\textcolor{cyan}{Ostapchenko, 2011}), EPOS-LHC (\textcolor{cyan}{Pierog et al., 2015}) and SIBYLL 2.3d (\textcolor{cyan}{Riehn et al., 2020}) to separate the KASCADE-Grande data into an electron-rich and electron-poor components dominated by H$+$He$+$C and Si$+$Fe nuclei, respectively. The results confirm the hardening and the softening at around $100$ PeV in the corresponding spectra of light and heavy elements. By summing the intensities of electron-rich and electron-poor components, the all-particle spectrum was reconstructed for QGSJET-II-04, SIBYLL 2.3d and EPOS-LHC in the  $10 \, \mbox{PeV} - 1 \, \mbox{EeV}$ interval. The second knee and the low-energy ankle are also observed in these results.  The knee-like feature of the heavy mass group is, however, observed at lower energies ($\sim 55$ PeV) in comparison with the spectra obtained with the pre-LHC hadronic interaction models.

     An additional confirmation of the above  features in the total, light and heavy spectra of cosmic rays come from a preliminary study on the electron and muon content of EAS performed with events reconstructed from the combined data of KASCADE and KASCADE-Grande in the full energy range from $1$ PeV to $1$ EeV (\textcolor{cyan}{Schoo et al., 2015}).  In a further analysis, KASCADE-Grande has  presented preliminary data that suggest  the existence of ankle-like features in the individual spectra of H and He above $10$ PeV (\textcolor{cyan}{Arteaga et al., 2019}).

     More recent analysis from ICETOP/ICECUBE, the Pierre Auger observatory, the Telescope array and the TAIGA complex have allowed to study in more detail the spectrum and composition of cosmic rays in the PeV energy range. These investigations have provided more evidence for the presence of the low-energy ankle and the second knee in the total spectrum, and have confirmed that composition becomes heavier towards $100$ PeV with different mass-composition parameters of EAS. In addition, the composition studies with the data from these experiments have shown that light primaries are more abundant around $1$ EeV. In (\textcolor{cyan}{Aartsen et al., 2019}), the all-particle  spectrum from $3$ PeV to $1$ EeV was measured independently with the ICETOP array and with the combined data of the ICETOP and ICECUBE detectors. The reconstructed spectra are in agreement with each other within systematic uncertainties. In this study, the knee is observed in the range of $5-6$ PeV, the low-energy ankle around $17$ PeV, while the second knee, at $\sim 160$ PeV. From the combined ICETOP/ICECUBE data, the energy spectra for the mass groups of H, He, O and Fe were also estimated in the interval $3 \, \mbox{PeV}  < E < 1 \, \mbox{EeV}$ (\textcolor{cyan}{Aartsen et al., 2019}). The statistics is not large enough yet to provide a detailed picture of the features in the elemental spectra, but allows to confirm that from the knee to the second knee, the composition evolves from light to heavy, and that close to $1$ EeV it tends to become lighter again, as shown also by the corresponding calculations of $\langle \ln A \rangle$.

     Recent upgrades at the Pierre Auger observatory have allowed to lower the energy threshold for the measurements of the total energy spectrum from the ultra-high energy regime up to $6$ PeV. Using the SD-750 array - with a surface of $27 \, \mbox{km}^2$, $63$ water Cherenkov detectors (WCD) and $750$ m spacing-, the Auger collaboration derived the spectrum at energies between $100$ PeV and $25$ EeV (\textcolor{cyan}{Abreu et al., 2021}). The second knee was observed, but due to its proximity to the energy threshold, it was not possible to measure the position of this structure. Although, it was determined that this feature is wider than the ankle. In another study,  it was possible to measure the total spectrum from $60$ PeV up to $6$ EeV with the SD-433 array, which is composed of $19$ WCD with $433$ m spacing spread over an area of $1.9 \, \mbox{km}^2$ (\textcolor{cyan}{Brichetto et al., 2023}). Here, the energy break for the second knee was found at $230$ PeV. Furthermore, in  (\textcolor{cyan}{Novotny et al., 2021}), the spectrum was estimated over more than two decades of energy, starting at $6$ PeV and ending around $1$ EeV, with the High Elevation Auger Telescopes (HEAT). Both the low-energy ankle and the second knee were observed at energies of $28$ and $158$ PeV, respectively. By using data from HEAT, the SD-750 and SD-1500 arrays, as well as hybrid data from the fluorescence telescopes and the SD-1500 detector, in  (\textcolor{cyan}{Novotny et al., 2021}), the Auger Collaboration derived a combined energy spectrum with high-statistics for all cosmic-ray primaries covering the energy interval from $6$ PeV up to $100$ EeV. The energy estimations, in general, were obtained from formulas calibrated with calorimetric measurements of EAS. An update of this work, for $E > 100$ PeV, was presented in (\textcolor{cyan}{Abreu et al., 2021}). 

     With regard to the composition measurements of cosmic rays in the PeV region, the Auger Collaboration has estimated the fractional abundances of the H, He, CNO and Fe nuclei in the flux of cosmic rays  (\textcolor{cyan}{Tkachenko et al., 2023}) and has provided data on $\langle X_{\mathrm{max}} \rangle$ (\textcolor{cyan}{Aab et al., 2017a}; \textcolor{cyan}{Mayotte et al., 2023}) above a few hundred PeV, which show that the mass of the primary particles is lighter around $1$ EeV than close to the second knee. Similar conclusions have been obtained with  preliminary analyses on the composition of cosmic rays performed with the Telescope Array Low energy Extension (TALE) (\textcolor{cyan}{Abu-Zayyad et al., 2023}; \textcolor{cyan}{Fujita et al., 2023}). 
     
     In (\textcolor{cyan}{Abu-Zayyad et al., 2023}), the TALE experiment, using data from the fluorescence detectors (FD) in monocular mode,  measured $\langle X_{\mathrm{max}} \rangle$ and the fractional abundances of the elemental groups of H, He, CNO and Fe primaries and the all-particle spectrum within the interval of $1 \, \mbox{PeV} - 1 \, \mbox{EeV}$. In this analysis, among other results, it was confirmed that the knee is dominated by light nuclei and that the composition becomes heavier from the knee to the second knee region. The mass composition of cosmic rays around the second knee was also investigated by TALE in an independent study using hybrid data (from the FD and the surface detectors) (\textcolor{cyan}{Fujita et al., 2023}). Here, $\langle X_{\mathrm{max}} \rangle$ was estimated as a function of the primary energy, and from the corresponding distributions, the fraction of H, CNO and Fe primaries  were obtained from $32$ PeV to $3.2$ EeV. The results are still preliminary but these data suggest a recovery in the abundances of the light mass group around $60$ PeV and also seem to imply that, inside the energy range under study, the elemental groups of H, CNO and Fe are dominant close to $1$ EeV, $60$ PeV  and  $200$ PeV, respectively. Regarding the measurements on the total spectrum of cosmic rays, TALE has recently presented a  combined spectrum for $2 \, \mbox{PeV} < E < 100 \, \mbox{EeV}$ (\textcolor{cyan}{Ivanov et al., 2021}), which was derived from the data of the low-energy extension, for energies between $2$ PeV and $2$ EeV, and measurements with the Telescope Array above $2$ EeV. In this spectrum, the knee is located at $\sim 3$ PeV, the low-energy ankle, close to $17$ PeV and the second knee, at around $110$ PeV.

     Finally, we will mention some updated measurements of the TAIGA complex. This experiment has studied the composition and spectrum of cosmic rays in different energy intervals that cover the PeV energy regime. With TAIGA-HiScore, the spectrum was measured recently around the knee as mentioned in the previous section (\textcolor{cyan}{Prosin et al., 2023}). Here, the knee was measured at $3$ PeV. Meanwhile, using the Tunka-133 array, the total spectrum was analyzed from $7$ PeV to $3$ EeV. The corresponding results showed the low-energy ankle at $20$ PeV and the second knee at around $300$ PeV (\textcolor{cyan}{Budnev et al., 2020}). Tunka-Grande has investigated the interval $9 \, \mbox{PeV} - 3  \, \mbox{EeV}$ and has found two spectral breaks, one at $20$ PeV and another one close to $100$ PeV, which corresponds to the low-energy ankle and the second knee, respectively (\textcolor{cyan}{Ivanova et al., 2023}). For the study of the cosmic-ray composition,  $\langle X_{\mathrm{max}} \rangle$ and $\langle \ln A \rangle$ have been investigated in the TAIGA complex. Recent results obtained with  combined  TAIGA-HiScore and Tunka-133 data are presented in (\textcolor{cyan}{Prosin et al., 2023a}). They also show an increment in the heavy component of cosmic rays in the region of the second knee. A compilation of the spectrum and composition measurements presented in this section are shown in Figs.~\ref{Allspec} - \ref{Fespectrum}. 
     \vspace{0.5pc}

\scriptsize 
\textbf{Acknowledgements.} I would like to acknowledge the organizers of the ECRS 2024 for the invitation to present this review and the financial support from CONAHCYT and the CIC-UMSNH. The data points were obtained from the original papers, the Cosmic Ray Data Base (\textcolor{cyan}{Maurin et al., 2023}) and the KCDC (\textcolor{cyan}{Haungs et al., 2018}).
\vspace{-1pc}

\section*{References}
\begin{list}{}%
{\leftmargin=1em \itemindent=-1em}
\scriptsize
\itemsep -2pt
\vspace{-0.5pc}
\item[] Aab, A. et al.: 2017, {\it Science} {\bf  357}, 1266.\vspace{-0.2pc}
\item[] Aab, A. et al.: 2017a, {\it  Phys. Rev. D} {\bf  96}, 122003.\vspace{-0.2pc}
\item[] Aab, A. et al.: 2020, {\it  Phys. Rev. Lett.} {\bf  125}, 121106.\vspace{-0.2pc}
\item[] Aartsen, M. G. et al.: 2019, {\it Phys. Rev. D} {\bf  100}, 082002.\vspace{-0.2pc}
\item[] Aartsen, M. G. et al.: 2020, {\it Phys. Rev. D} {\bf  102}, 122001.\vspace{-0.2pc}
\item[] Abbasi, R.U. et al.: 2008, {\it  Phys. Rev. Lett.} {\bf  100}, 101101.\vspace{-0.2pc}
\item[] Abbasi, R.U. et al.: 2018, {\it  Astrophys. J.} {\bf  865}, 74.\vspace{-0.2pc}
\item[] Abraham, J. et al.: 2008, {\it  Phys. Rev. Lett.} {\bf  101}, 061101.\vspace{-0.2pc}
\item[] Abreu, P. et al.: 2021, {\it Eur. Phys. J. C}  {\bf 81}, 966.  \vspace{-0.2pc}
\item[] Abdul Halim, A. et al.: 2024, {\it J. Cosmol. Astropart. Phys.}   {\bf 01}, 022.  \vspace{-0.2pc}
\item[] Abu-Zayyad, T. et al.: 2002, astro-ph/0208243.  \vspace{-0.2pc}
\item[] Abu-Zayyad, T. et al.: 2023, {\it  PoS (ICRC2023)}  {\bf 379}.  \vspace{-0.2pc}
\item[] Adriani, O. et al.: 2019, {\it  Phys. Rev. Lett.}  {\bf 122}, 181102. \vspace{-0.2pc}
\item[] Adriani, O. et al.: 2023, {\it  Phys. Rev. Lett.}  {\bf 130}, 171002. \vspace{-0.2pc}
\item[] Albert, A. et al.: 2022, {\it Phys. Rev. D}  {\bf 105}, 063021.  \vspace{-0.2pc}
\item[] Alemanno, F. et al.: 2021, {\it Phys. Rev. Lett.}  {\bf 126},  201102 .  \vspace{-0.2pc}
\item[] Alemanno, F. et al.: 2024, {\it Phys. Rev. D}  {\bf 109},  L121101 .  \vspace{-0.2pc}
\item[] Alfaro, R. et al.: 2017, {\it Phys. Rev. D}  {\bf 96}, 122001.  \vspace{-0.2pc}
\item[] Alfaro, R. et al.: 2025, {\it Astrop. Phys.}  {\bf 167}, 103077.  \vspace{-0.2pc}
\item[] An Q.,et al.: 2019, {\it Sci. Adv.}  {\bf 5}, eaax3793. \vspace{-0.2pc}
\item[] Antoni, T. et al.: 2005, {\it Astrop. Phys.}  {\bf 24}, 1.  \vspace{-0.2pc}
\item[] Apel, W.D. et al.: 2009, {\it Astrop. Phys.}  {\bf 31}, 86.  \vspace{-0.2pc}
\item[] Apel, W.D. et al.: 2011, {\it Phys. Rev. Lett.}  {\bf 107}, 171104.  \vspace{-0.2pc}
\item[] Apel, W.D. et al.: 2012, {\it Astrop. Phys.}  {\bf 36}, 183.  \vspace{-0.2pc}
\item[] Apel, W.D. et al.: 2013, {\it Astrop. Phys.}  {\bf 47}, 54.  \vspace{-0.2pc}
\item[] Apel, W.D. et al.: 2013a, {\it Phys. Rev. D}  {\bf 87}, 081101(R).  \vspace{-0.2pc}
\item[] Arteaga-Velázquez, J.C. et al.: 2015, {\it J. Phys.: Conf. Ser.}  {\bf 651}, 012001.\vspace{-0.2pc}
\item[] Arteaga-Velázquez, J.C. et al.: 2019, {\it EPJ Web Conf.}  {\bf 6003}.\vspace{-0.2pc}
\item[] Arteaga-Velázquez, J.C. et al.: 2023, {\it PoS (ICRC2023)}  {\bf 299}.\vspace{-0.2pc}
\item[] Atkin, E. et al.: 2017, {\it J. Cosmol. Astropart. Phys.}   {\bf 07}, 020.  \vspace{-0.2pc}
\item[] Auger, P. et al.: 1939, {\it Rev.  Mod. Phys.}   {\bf 11}, 288.  \vspace{-0.2pc}
\item[] Bahcall, J. N. and Waxman, E.: 2002, arXiv:hep-ph/0206217.  \vspace{-0.2pc}
\item[] Barret, P. H. et al.: 1952, {\it Rev. Mod. Phys.}  {\bf 24},  133. \vspace{-0.2pc}
\item[] Bergman, D.R.; Belz, J.W.: 2007, {\it J. Phys. G: Nucl. Part. Phys.}  {\bf 34},  R359. \vspace{-0.2pc}
\item[] Bird, D.J. et al.: 1993,  {\it Phys. Rev. Lett.}  {\bf 71}, 3401. \vspace{-0.2pc}
\item[] Bird, D.J. et al.: 1994,  {\it Astrophys. J.}  {\bf 424}, 491. \vspace{-0.2pc}
\item[] Bird, D.J. et al.: 1995,  {\it Astrophys. J.}  {\bf 441}, 144. \vspace{-0.2pc}
\item[] Bowen, I.S. et al.: 1938, {\it  Phys. Rev.}  {\bf 53}, 855.  \vspace{-0.2pc}
\item[] Budnev, N.M. et al.: 2020, {\it  Astrop. Phys.}  {\bf 117}, 102406.  \vspace{-0.2pc}
\item[] Brichetto, A. et al.: 2023, {\it  PoS (ICRC2023)}  {\bf 398}.  \vspace{-0.2pc}
\item[] Castellina, A. et al.: 2023,  {\it  SciPost Phys. Proc.}  {\bf 13}, 034.  \vspace{-0.2pc}
\item[] Cao, Z., et al.: 2024, {\it  Phys. Rev. Lett.} {\bf  132}, 131002.\vspace{-0.2pc}
\item[] Choi, G.H. et al.: 2022, {\it Astrophys. J.} {\bf 940}, 107. \vspace{-0.2pc}
\item[] Clark, G. et al.: 1957,  {\it Nature} {\bf 180}, 4583.\vspace{-0.2pc}
\item[] Comisso, L. et al.: 2024,  {\it Astrophys. J. Lett.} {\bf 977}, L18.\vspace{-0.2pc}
\item[] Crawshaw, J. K. et al.: 1956, {\it Proc. Phys. Soc. Sec. A} {\bf 69(2)}, 102. \vspace{-0.2pc}
\item[] Cunningham, C. et al.: 1980, {\it Astrophys. J.} {\bf 236}, L71. \vspace{-0.2pc}
\item[] Efimov, N.N. et al.: 1991, {\it Proc. ICRR Int. Symp.}, M. Nagano, and F. Takahara (eds.),  World Scientific, Singapore, 20. \vspace{-0.2pc}
\item[] Fuhrmann, D.: 2012, {\it PhD thesis}, University of Wuppertal. \vspace{-0.2pc}
\item[] Fujita, K.: 2023,  {\it PoS (ICRC2023)} {\bf 401}.\vspace{-0.2pc}
\item[] Giacinti, G.; Semikoz, D.: 2023,  arXiv:2305.10251 [astro-ph.HE].\vspace{-0.2pc}
\item[] Grebenyuk, V.  et al.: 2019, {\it Adv. Sp. Res. } {\bf 64}, 2546. \vspace{-0.2pc}
\item[] Hess, V.F.: 1912, {\it Phys. Z.}   {\bf 13}, 1084.\vspace{-0.2pc}
\item[] Haungs, A. et al: 2003, {\it Rep. Prog. Phys.}   {\bf 66(7)}, 1145.\vspace{-0.2pc}
\item[] Haungs, A. et al: 2018, {\it Eur. Phys. J. C}  {\bf 78}, 741.\vspace{-0.2pc}
\item[] Ivanov, D.: 2021,  {\it PoS (ICRC2021)} {\bf 341}.\vspace{-0.2pc}
\item[] Ivanova, A. L.: 2023,  {\it SciPost Phys. Proc.} {\bf 13} 011.\vspace{-0.2pc}
\item[] Kachelriess, M.: 2019, {\it EPJ Web Conf.} {\bf 210}, 04003.\vspace{-0.2pc}
\item[] Kang, D.: 2023, {\it PoS (ICRC2023)} {\bf 307}.\vspace{-0.2pc}
\item[] Kulikov, G.V. and  Khristiansen G.B.: 1958, {\it ZhETF} {\bf 35}, 635.\vspace{-0.2pc}
\item[] Lawrence, M.A.: 1991,   {\it  J. Phys. G: Nucl. Part. Phys.}  {\bf  17}, 733.\vspace{-0.2pc}
\item[] Linsley, J.: 1963,   {\it Phys. Rev. Lett.}  {\bf  10}, 146.\vspace{-0.2pc}
\item[] Linsley, J.: 1963a,  {\it  Proc 8th ICRC} (Jaipur)  {\bf  4}, 77.\vspace{-0.2pc}
\item[] Maestro, P.: 2015, {\it PoS (ICRC2015)}  {\bf 016}.\vspace{-0.2pc}
\item[] Maurin, D.: 2023, {\it Eur. Phys. J. C}  {\bf 83}, 971.\vspace{-0.2pc}
\item[] Mayotte, E.: 2023, {\it PoS (ICRC2023)}  {\bf 365}.\vspace{-0.2pc}
\item[] Mollerach, S. et al.: 2018, {\it Prog. Part. Nuc. Phys.}  {\bf 98}, 85.\vspace{-0.2pc}
\item[] Montini, P. et al.: 2016, {\it Nuc. Part. Phys. Proc.}  {\bf 279}, 7.\vspace{-0.2pc}
\item[] Nagano, M. et al.: 1992, {\it J. Phys. G: Nucl. Part. Phys.}  {\bf 18}, 423.\vspace{-0.2pc} 
\item[] Navas, S. et al.: 2024, (Particle Data Group) {\it Phys. Rev. D} {\bf 110}, 030001. \vspace{-0.2pc}
\item[] Novotny, V. et al.: 2021,  {\it PoS (ICRC2021)} {\bf 324}.\vspace{-0.2pc}
\item[] Ostapchenko, S.: 2006,  {\it Phys. Rev. D} {\bf 74}, 014026.\vspace{-0.2pc}
\item[] Ostapchenko, S.: 2011,  {\it Phys. Rev. D} {\bf 83}, 014018.\vspace{-0.2pc}
\item[] Ostrowski, M.: 2002,  {\it 	J. Phys. Stud.} {\bf 6}, 393.\vspace{-0.2pc}
\item[] Panov, A.D. et al.: 2009,  {\it Bull. Russ. Acad. Sci. Phys.} {\bf 73}, 564.\vspace{-0.2pc}
\item[] Peters, B.: 1961,  {\it Nuovo Cimento} {\bf 22}, 800.\vspace{-0.2pc}
\item[] Pierog, T. et al.: 2015,  {\it Phys. Rev. C} {\bf 92}, 034906.\vspace{-0.2pc}
\item[] Prosin, W. et al.: 2023,  {\it Bull. Russ. Acad. Sci. Phys.} {\bf 87}, 1043.\vspace{-0.2pc}
\item[] Prosin, W. et al.: 2023a,  {\it SciPost Phys. Proc.} {\bf 13}, 037.\vspace{-0.2pc}
\item[] Riehn, F. et al.: 2020,  {\it Phys. Rev. D} {\bf 102}, 063002.\vspace{-0.2pc}
\item[] Schoo, S. et al.: 2015,  {\it PoS (ICRC2015)} {\bf 263}.\vspace{-0.2pc}
\item[] Tkachenko, O. et al.: 2023,  {\it PoS (ICRC2023)} {\bf 438}.\vspace{-0.2pc}
\item[] Varsi, F. et al.: 2021, {\it PoS (ICRC2021)}  {\bf 388}.\vspace{-0.2pc}
\item[] Varsi, F. et al.: 2024, {\it  Phys. Rev. Lett.} {\bf  132}, 051002.\vspace{-0.2pc}
\end{list}

\end{document}